\documentclass[aps,prx,amsmath,amssymb,reprint,floatfix]{revtex4-2}
\usepackage{graphicx}
\usepackage{bm}
\usepackage{dcolumn}
\usepackage{hyperref}
\usepackage{graphicx}
\graphicspath{{images/}}
\usepackage{dcolumn}
\usepackage[utf8]{inputenc}
\usepackage{amssymb}
\usepackage[dvipsnames]{xcolor}
\usepackage[normalem]{ulem}
\usepackage{amsmath}
\usepackage{bm}
\usepackage{float}

\begin{document}


\title{Squeezing measurement with spectrally broadened in microstructured fiber local oscillator}


\author{A. Chudakov$^{1,2}$}
\email{chudakov.as@phystech.edu}

\author{V. Severin$^1$}

\author{A. Poshevkina$^{1,2}$}

\author{A. Finochenko$^{2,3}$}

\author{K. Kuznetsov$^{1,3}$}

\author{A. Levchenko$^{4}$}

\author{E. Serebryannikov$^{1,3}$}

\author{A. Fedotov$^{1,3}$}

\author{D. Kalashnikov$^{1,3}$}
\email{d.kalashnikov@rqc.ru}

\author{P. R. Sharapova$^{1,3}$}
\email{p.sharapova@rqc.ru}
 

\affiliation{$^{1}$Russian Quantum Center, Bolshoy Bulvar 30, bld. 1, 121205 Moscow, Russia}

\affiliation{$^{2}$Moscow Institute of Physics and Technology, 141700 Dolgoprudny, Russia}

\affiliation{$^{3}$Department of Physics, M.V.Lomonosov Moscow State University, Leninskie Gory, 119991 Moscow, Russia}

\affiliation{$^{4}$Dianov Fiber Optics Research Center, Prokhorov General Physics Institute of RAS, 119333 Moscow, Russia}


\begin{abstract}
Recently, continuous variables (CV) quantum computing, i.e. computing employing squeezed states, has attracted significant attention,  as it enables to increase the dimensionality of the system. In CV protocols, homodyne detection plays a key role in the realization of the required gates and operations.
Multimode squeezed light generated via parametric down-conversion (PDC) benefits from the broadband bandwidth, which encompasses multiple frequency modes. However, its broadband structure simultaneously requires the precise matching of the local oscillator (LO) with the squeezed light modes for their correct homodyne detection and processing. Achieving such matching is usually a challenging task, since the spectral bandwidth of the local oscillator is typically significantly narrower than the bandwidth of the squeezed-light modes. In this work, we propose and experimentally demonstrate a technique for expanding the bandwidth of the LO by passing it through the microstructured fiber. We validate the proposed concept by measuring squeezing in the modes of multimode squeezed light generated in type-0 and type-II PDC. We show that the broadened LO starts to spectrally match the modes generated via type-II PDC. The presented technique might be beneficial for implementing various CV systems and protocols.
\end{abstract}


\maketitle

\section{Introduction}

Squeezed states of light are characterized by reduced noise uncertainty in a specific quadrature of the electromagnetic field compared to fundamental shot noise. This aspect makes them an essential tool for a number of practical applications ranging from quantum metrology and sensing to quantum communications and computations \cite{doi:10.1126/science.1104149, PhysRevLett.110.181101, doi:10.1021/acsphotonics.9b00250, PhysRevA.109.053715, doi:10.1126/science.282.5389.706, Nat_Com_Madsen2012, PhysRevLett.97.110501, doi:10.1126/science.aay2645, doi:10.1126/science.aay4354}. Among this diverse family of possible implementations, applications utilizing squeezed light with multiple time-frequency and spatial modes (multimode squeezed light) have become of particular interest in recent years. Indeed, they lie at the heart of the concept of measurement based quantum computing (MBQC), which is one of the prominent ways of realizing fault-tolerant quantum computations in continuous variables (CV),  additionally benefiting in a significantly expanded Hilbert space 
\cite{PhysRevA.64.012310, PhysRevLett.112.120504, NatCom.8.15645, AghaeeRad2025, Roh2025, malik2026}. 
One of the popular schemes for generating multimode squeezed light is based upon parametric down-conversion (PDC) in nonlinear media using an ultrafast pulsed pump, which provides direct access to a multitude of spectral and temporal modes. In this context, the synchronously pumped optical parametric oscillator (SPOPO) has proven to be a versatile tool for generating multimode squeezed states \cite{PhysRevLett.108.083601, NatCom.8.15645, Roh2025, Gwak26}. However, the SPOPO supermodes are highly demanding in terms of pump properties and intracavity dynamics, which is not easily adjustable in the experiment \cite{Thorpe2008, PhysRevLett.124.163601, Averchenko_2024, 10.1063/5.0156331, Suerra2026}. At the same time, single-pass optical parametric amplifiers (OPAs) constitute a much simpler, cavity-free experimental alternative. 
In the case of OPA, the generated PDC radiation is characterized by the THz-order bandwidth, which means that multiple frequency modes are produced in one pass, ensuring reconfigurable entanglement links and mode-selective non-Gaussian operations \cite{NatCom.8.15645, NatPhys.16.144}. However, when using a multimode light, attention should be paid to its correct measurement procedure.
 
The standard approach to measuring squeezed states of light is based on homodyne detection, where the light under study is superimposed with a strong phase-coherent reference beam, the so-called local oscillator (LO), followed by balanced detection \cite{PhysRevA.13.2226, RevModPhys.81.299}. The LO and pump for generating squeezed light are usually prepared from the same laser source. This results in the important caveat: The LO bandwidth becomes significantly narrower than the bandwidth of the squeezed-light modes generated in OPA.  However, for accurate measurement, the LO  must perfectly match the generated squeezed-light modes  \cite{10.1063/5.0156331, PhysRevResearch.6.043113}. Decreasing the pulse duration and, correspondingly, increasing the bandwidth of the pump allows to broaden the LO. But this also reduces the pump coherence length and relaxes the PDC phase-matching conditions. As a result, squeezed light with an even broader spectrum is generated, and the modes under study once again become spectrally mismatched with the local oscillator.
Furthermore, the use of shorter pump pulses generally complicates the experimental setup, and  additional care should be taken to mitigate group velocity dispersion effects.

 In this work, we present a technique to spectrally broaden the bandwidth of the LO by passing it through a microstructured fiber and matching it with the PDC frequency modes for  squeezing measurements.
 Microstructured fibers are commercially available products which are widely used nowadays for a number of practical applications ranging from biomedicine and optical communications to precision measurements and sensing \cite{Lee_2012, 4137601, Schibli2008, Wang:22}. In addition, they were actively used for spectral engineering such as supercontinuum generation and pulse compression \cite{853507, PhysRevLett.87.203901, Genty:04, Mollenauer:83, Travers:07, Travers2019}. In this work, the self-phase modulation effect of a microstructured fiber is utilized to spectrally match the LO with the modes of the generated squeezed light.
 The degree of broadening can be controlled by the number of parameters related to the pump (pump power, pump duration, pump polarization), the fiber itself (fiber length, zero dispersion wavelength) and coupling strength.   However, together with broadening, the LO gets a spectral chirp during its propagation in the microstructured fiber. The successful implementation of the broadened LO for squeezing measurements also requires compensation for such a chirp.  We show that the broadened LO provides the level of squeezing comparable to that before the procedure, but benefits in the expanded bandwidth. The presented technique provides greater flexibility in the LO shaping and might be useful in the experimental realizations of MBQC schemes based on single-pass OPA.

 \section{Theory}

In this work, the PDC process in the single-pass OPA is considered collinear and frequency-degenerate, it can be described by the effective Hamiltonian \cite{PhysRevA.97.053827}:
\begin{equation}
\hat{H} = \Gamma \sum_{n} \sqrt{\lambda_{n}} \, \hat{A}_{n}^{\dagger} \hat{B}_{n}^{\dagger}+ \mathrm{h.c.},
\label{eq: Hamiltonian}
\end{equation}
where $\Gamma$ is the coupling constant depending on the pump and nonlinear crystal parameters,  $\hat{A}_{n}^\dagger$ and $\hat{B}_{n}^\dagger$  are the broadband creation operators (Schmidt operators) related to the spectral broadband modes (supermodes) $u_n(\omega_s)$ and $v_n(\omega_i)$, respectively, while $\omega_{s,i}$ are the frequencies of the signal and idler photons.  The broadband modes are obtained via the Schmidt decomposition of the joint spectral amplitude (JSA):
\begin{equation}
F(\omega_s, \omega_i) = \sum_n  \sqrt{\lambda_n} \, u_n(\omega_s)\, v_n(\omega_i),
\label{eq: JSA6}
\end{equation}
where  $\lambda_n$ are the Schmidt eigenvalues characterizing the degree of squeezing in each broadband mode. 
For an unchirped pump with a Gaussian temporal envelope and central frequency $\omega_p$ the JSA  can be written as \cite{Ebers_2022} 

\begin{eqnarray}
F(\omega_s, \omega_i) & = & \frac{1}{N} \exp\left(-\frac{(\omega_p - \omega_s - \omega_i)^2 \tau^2}{2}\right) \nonumber\\
&&\times\mathrm{sinc}\left(\frac{\Delta k L}{2}\right) \exp\left[i \frac{\Delta k L}{2}\right],
\label{eq: JSA1}
\end{eqnarray}
where $N$ is the normalization constant, $\tau$ is related to the pump pulse duration $ \tau_p$ as  $ \tau_p= 2\sqrt{ln2}\tau$, and $L$  is the crystal length. The phase-mismatch is defined as $\Delta \vec{k} = \vec{k}_p - \vec{k}_s - \vec{k}_i - \frac{2\pi}{\Lambda}=0$, with  $\vec{k}_{s,i,p}$  denoting the wave vectors of the pump, signal, and idler fields, respectively, and $\Lambda$ is the poling period of the crystal.
Assuming an unchirped gaussian pump, the Schmidt modes can be approximated by the Hermite-Gaussian (HG) functions \cite{Patera2010}. The signal (and similarly for idler) spectral intensity distribution  can be expressed via the broadband modes as follows  \cite{PhysRevA.97.053827}:
\begin{equation}
I(\omega_s) = \sum_n \left|u_n(\omega_s)\right|^2 \mathrm{sinh}^2 \left[ \sqrt{\lambda_n}  \Gamma \  T \right], 
\label{eq: Theor spectra}
\end{equation}
where $T$ is the interaction time. In the low gain-regime, it coincides with integrating the JSA over the conjugate frequency variable:
$\displaystyle\int \left|F(\omega_s,\omega_i)\right|^2 d\omega_i $.

An arbitrary spectral LO profile  can be decomposed with respect to the broadband PDC modes, that form a basis, as: 
\begin{equation}
\psi_{LO} =\sum_n M_{n} \, e^{i\theta_{n}} u_{n}(\omega),
\label{eq: LO_arb}
\end{equation}
where  $\theta_{n}$ and $M_{n}$ are the phase and amplitude of the expansion coefficient, respectively, with $\sum_n M_{n}^2=1$. 
In the balanced homodyne detection scheme for type-0 or type-I PDC (where the signal and idler photons have identical polarizations), the variance of the photocurrent difference reads \cite{PhysRevA.73.063819}
\begin{equation}
\left\langle\delta I_{c}\right\rangle \sim\sum_{n}\frac{M_{n}^2}{4}(e^{2r_{n}}\sin^{2}{(\theta_{n}+\phi)}+e^{-2r_{n}}\cos^{2}{(\theta_{n}+\phi)}),
\label{eq: Quadrature}
\end{equation}
where $r_{n}= \sqrt{\lambda_n}  \Gamma \ T $ is the level of squeezing in the $n$-th broadband mode, $\phi$ is the LO phase.

In the case of type-II PDC (where the signal and idler photons have orthogonal polarizations), the second local oscillator $\varphi_{\mathrm{LO}}(\omega) = \sum_{n} C_n e^{i\chi_n} v_n(\omega)$ with $\sum_n  C_n^2 = 1$ is required to measure the bipartite  squeezing.

In this case, the variance of the photocurrent difference can be expressed as \cite{chudakov2026}: 
\begin{eqnarray}
\left\langle\delta I_{c}\right\rangle & \sim \sum_{n}\big[ e^{2r_n} \big(M_n^2 + C_n^2 - 2 M_n C_n \cos{(\theta_n + \chi_n +2\phi)}\big)
\nonumber\\
& + e^{-2r_n} \big(M_n^2 + C_n^2 + 2 M_n C_n \cos{(\theta_n + \chi_n+2\phi)}\big) \big], 
\label{eq: Q_var type-2}
\end{eqnarray}
where we supposed the same phase $\phi$ for both LOs.

As the LO propagates through a microstructured fiber, its spectral and phase profiles undergo significant changes. The parameters of the considered microstructered fiber and the evolution of the pulse within the propagation are described in the Appendix \ref{app: Pulse-propagation}.

\section{Experiment}

\begin{figure}[htbp]
\centering\includegraphics[width=0.5\textwidth]{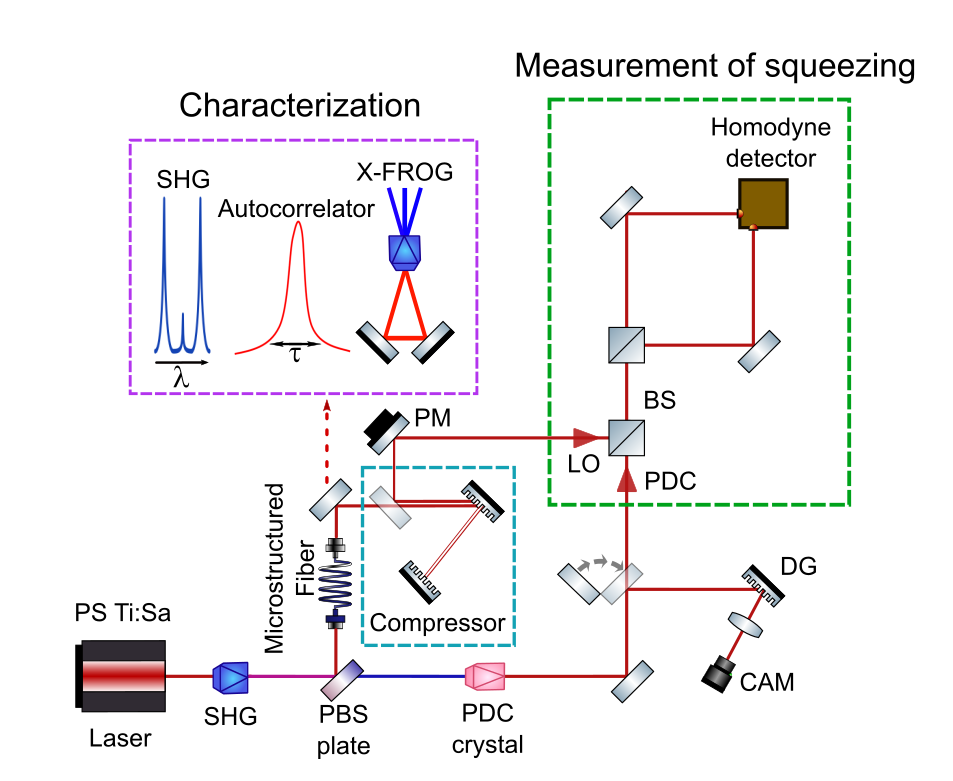}
\caption{Experimental setup: Laser - picosecond Ti:Sapphire Coherent Mira 900D; SHG - generation of the second harmonic in the LBO crystal; PBS plate separates the second harmonic from the LO, the LO is then passed through the microstructured fiber; PDC crystal - either 1 mm PPKTP type-0 or 2 mm PPKTP type-II; CAM - camera; DG - diffraction grating; PM - piezo mirror; BS - beamsplitter cube}
\label{setup}
\end{figure}

We performed the experiment using the setup shown in Fig. \ref{setup}, where the picosecond Ti:Sapphire Coherent Mira 900D pulsed laser was used as the fundamental pump with an average output power of $0.9\ \text{W}$, the pulse repetition rate of approximately $76\ \text{MHz}$, the duration of about $2\ \text{ps}$ and with the central wavelength of about $783.3\ \text{nm}$. The fundamental pump was used to produce the second-harmonic generation (SHG, type-I) at the wavelength of $391.65\ \text{nm}$  in the lithium triborate (LBO) crystal of the length of $17\ \text{mm}$. The polarizing beam-splitting plate (PBS-plate) separates the remaining fundamental pump from the SHG (here and after referred to as the pump); the remaining fundamental pump was later used as the LO. The UV pump with an average power of about $40\ \text{mW}$ passed through the periodically poled potassium titanyl phosphate (PPKTP) crystal with the length of $1\ \text{mm}$ for type-0 (Raicol, polling period $2.95\ \mu m$) and $2\ \text{mm}$ for type-II (Raicol, polling period $7.95\ \mu m$) PDC (PDC crystal), resulting in the generation of the multimode squeezed light. The PDC spectra for both type-0 and type-II were characterized by the self-made spectrometer consisting of the diffraction grating (Thorlabs GR-1208), the focusing lens, and the camera CAM (Thorlabs CS-135).

After separation from SHG, the LO was passed through the microstructered fiber of length  $21 cm$, with a mode diameter of 2 $\mu m$ and a zero-dispersion wavelength of 760 nm (produced by Fiber Optics Research Center of RAS), where it experienced spectral broadening due to self-phase modulation. The LO was coupled to the fiber with 40x objective, NA=0.4, and the similar objective collimated the beam after the fiber. The overall efficiency constituted around 20 $\% $. The power of the LO before the fiber and, correspondingly, the degree of broadening were controlled by the polarizing beamsplitter and half-wave plate (not shown in  Fig. \ref{setup}). The spectral profile of the LO before and after propagation through the microstructured fiber was measured by the optical spectrum analyzer (Thorlabs OSA 202C). After the microstructerd fiber the spectrally broadened LO was sent to the self-made compressor consisting of a pair of diffractive gratings (Thorlabs GR25-1208, blaze wavelength 750 nm, 1200 grooves/mm). The accuracy of the chirp compensation was controlled by SHG, pulse duration, and X-FROG measurements (for details see Appendix \ref{app: Chirp-control}).

Then, to provide squeezing measurements for type-0 PDC, the generated multimode squeezed light was recombined with the LO at the beam splitter BS and the level of squeezing was measured in the standard homodyne detection scheme by the self-made balanced detector (estimated efficiency 70 $\% $). For type-II PDC, the measuring block was modified, for details see  \cite{chudakov2026}.  We performed squeezing measurements using different spectral profiles of the broadened LO with and without chirp compensation.
All measurements were performed at the same LO power level of 4.5 mW. The relative phase $\phi$ between the LO and the squeezed light was swept using the piezo mirror PM introduced into the optical path of the LO, which was driven by the sawtooth signal (peak voltage $20\ \text{V}$, frequency $1\ \text{Hz}$). The measured photocurrent difference  was digitized using the Analog-to-digital converter (Agilent U1084A), the collected signal was processed by the developed software. Finally, the quadrature variance was calculated from the measured statistics. 

\section{Results and Discussion}

\begin{figure}[b]
\centering
\includegraphics[width=0.5\textwidth]{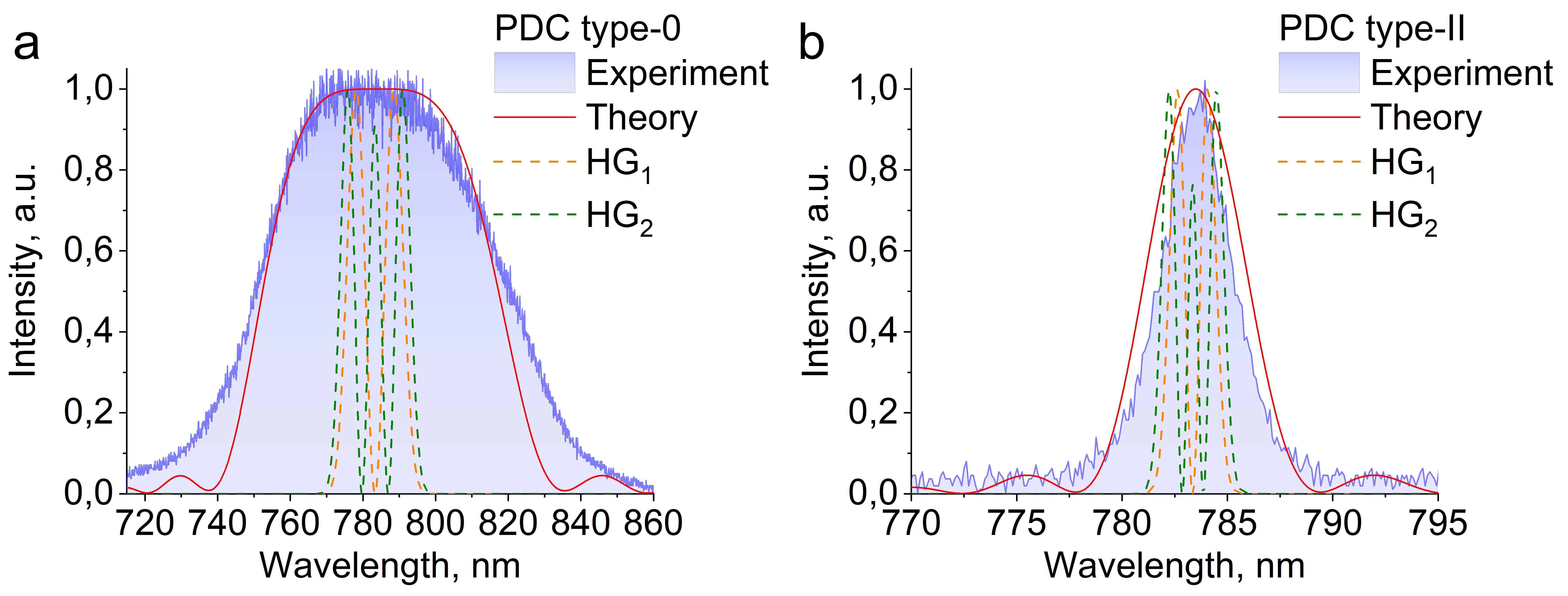}
\caption{Spectral distributions: a) type-0 PDC, b) type-II PDC. Experimentally measured spectra (shaded blue area) are shown together with theoretically calculated spectra (red solid lines). Theoretically calculated intensity distributions for the first and the second Schmidt modes (that correspond to the first and the second Hermite-Gaussian modes $HG_{1}$ and $HG_{2}$, respectively) are shown by the dashed orange and the dashed green lines, respectively. }
\label{fig: Spectral distributions}
\end{figure}

\begin{figure}[h]
\centering\includegraphics[width=0.5\textwidth]{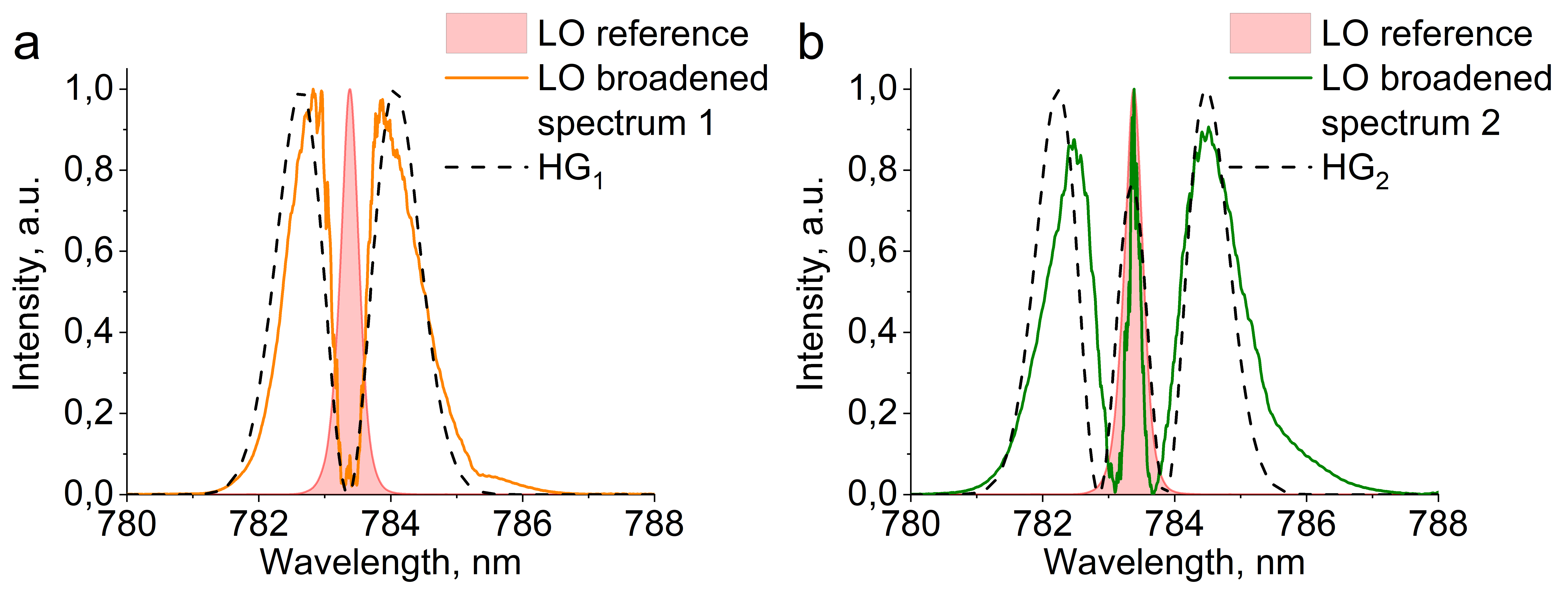}
\caption{LO spectra.  a) Spectrum 1. The  broadened LO spectrum (orange solid line) compared to the $HG_{1}$ type-II PDC mode (black dashed line). b) Spectrum 2. The  broadened LO spectrum (green solid line) compared to the $HG_{2}$ type-II PDC mode (black dashed line). The reference LO (before passing through the microstructured fiber) is shown by the red shaded area for both cases.}
\label{fig: Spectra LO}
\end{figure}

Fig. \ref{fig: Spectral distributions} shows the examples of the supermode distributions together with the spectra for type-0 and type-II PDC calculated according to Eq. (\ref{eq: JSA6} - \ref{eq: Theor spectra}) using the experimental parameters mentioned above. Note that in our case of low gain, Eq. (\ref{eq: Theor spectra}) coincides with integrating the JSA over the conjugate frequency variable.  The measured spectra are presented in the same figure as blue shaded areas. In both cases, the spectral width of the modes significantly exceeds the linewidth of the reference LO (0.33 nm) shown in Fig. \ref{fig: Spectra LO}, i.e. the emission of the Ti:Sapphire laser before passing through the microstructured fiber. As it propagates through the fiber, the LO undergoes broadening and modulation depending on the input power  due to the self-phase modulation effect. The measured LO spectra after broadening are presented in Fig. \ref{fig: Spectra LO}. Their widths are now becoming comparable to the characteristic spectral widths of squeezed-light modes, particularly in the type-II PDC case, see their intensity distributions indicated by black dashed lines in Fig. \ref{fig: Spectra LO}. 

However, the process of LO spectral broadening in the microstructured fiber induces an undesired chirp to the LO. This chirp was compensated using the diffraction-grating-based pulse compressor, where by varying the distance between the gratings, the long-wave LO spectral components were delayed with respect to the short-wave ones, thereby ensuring their temporal overlap and phase alignment. To verify that we implement a proper chirp compensation, we performed the SHG, pulse duration and X-FROG measurements for the broadened LO before and after such compensation (see Appendix \ref{app: Chirp-control}). 

After broadening and chirp compensation, the spectral width and shape of the LO closely resemble the spectral parameters of the Schmidt modes, especially for  type-II PDC, see Fig. \ref{fig: Spectra LO}, while the phase becomes more flat (see Fig. \ref{fig: X-FROG} in Appendix \ref{app: Chirp-control}). Implementing an additional well-known  pulse-shaping technique  using a spatial light modulator \cite{10.1063/5.0156331, chudakov2026, Monmayrant_2010} would allow to further improve the degree of similarity and, thereby, completely mimic the spectral-phase profiles of the PDC modes.  

\begin{figure*}
\centering\includegraphics[width=0.7\textwidth]{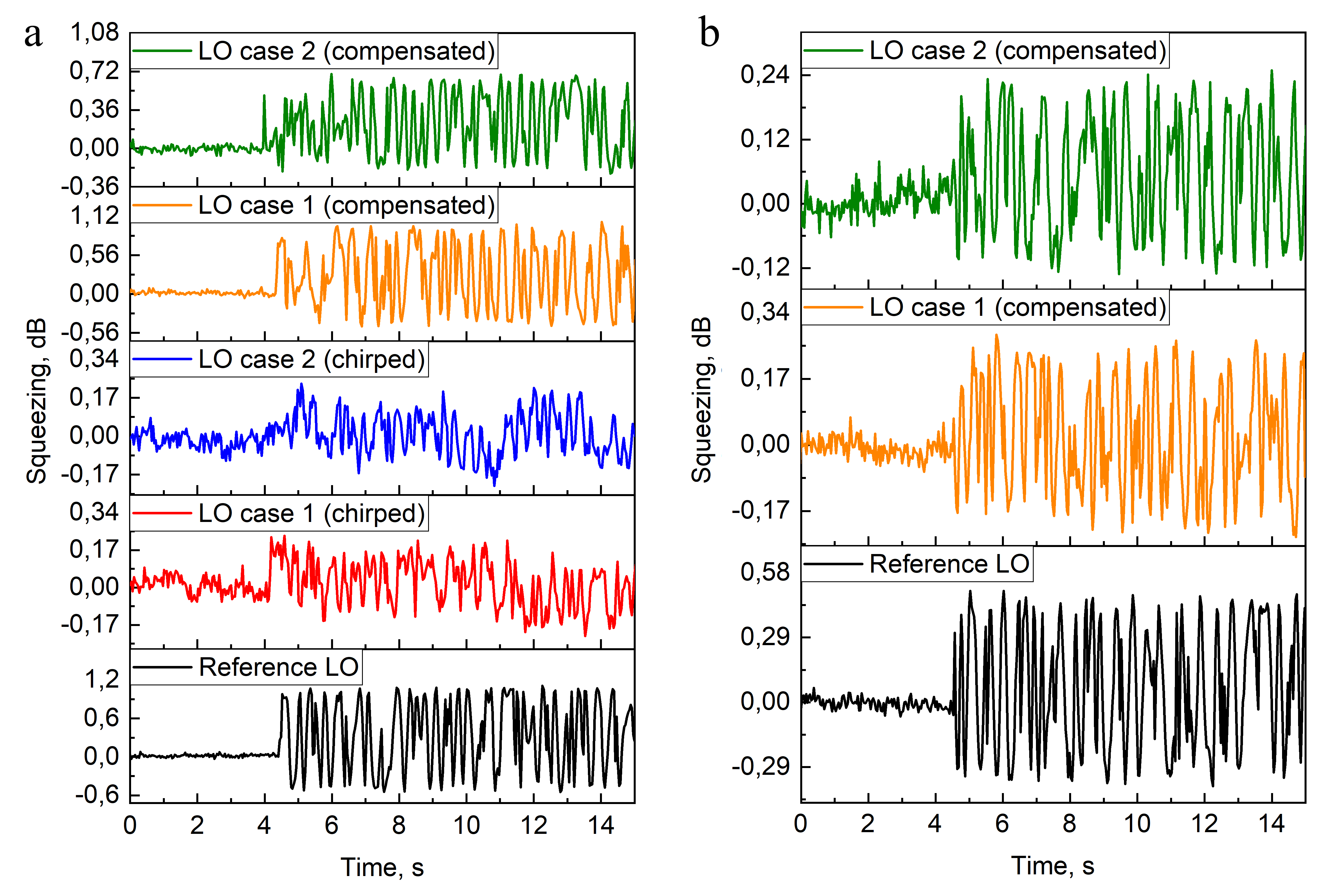}
\caption{Squeezing measurements for a) type-0  and b) type-II PDC. Black color corresponds to measurements with the reference (unbroadened) LO. Red and blue colors refer to  measurements with the broadened and chirped LOs, while orange and green colors - to the broadened and chirp-compensated LOs. Case 1 corresponds to spectrum 1, case 2  - to spectrum 2 in Fig. \ref{fig: Spectra LO}.}
\label{fig: Squeezing}
\end{figure*}

\begin{figure}
\centering\includegraphics[width=0.5\textwidth]{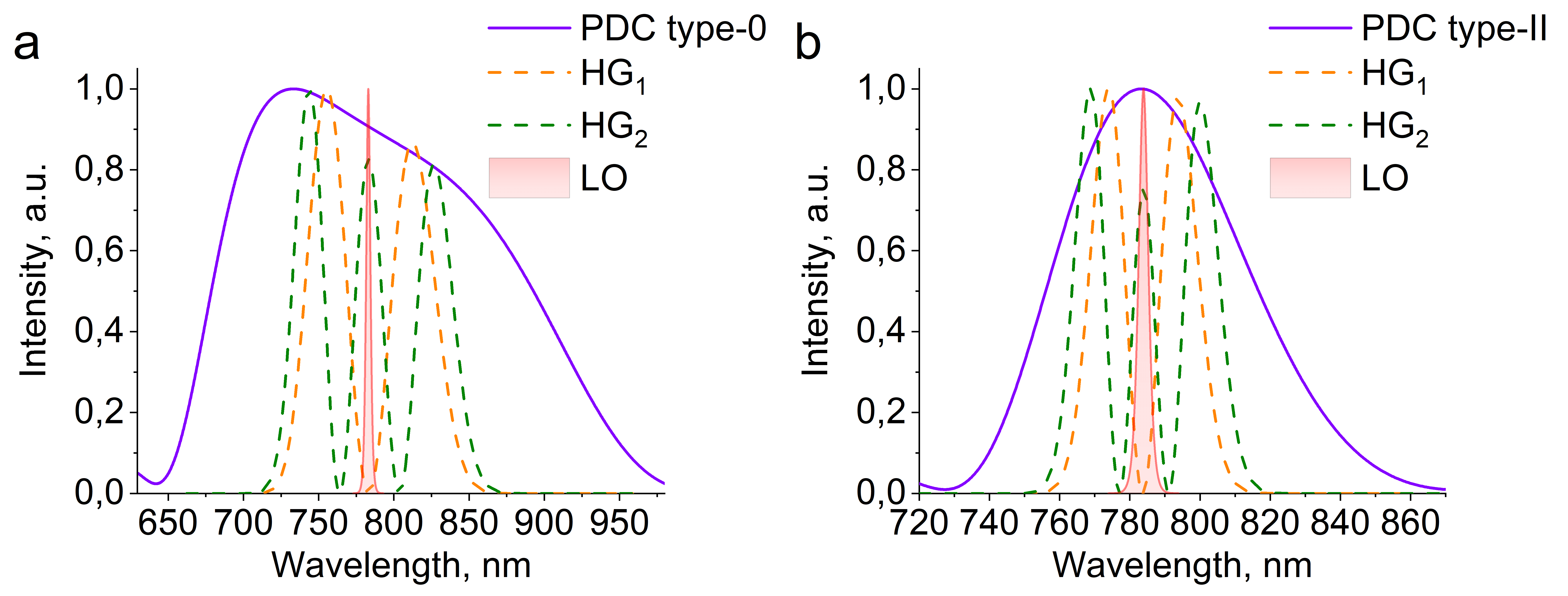}
\caption{Calculated modes and spectra for a) type-0 PDC and b) type-II PDC assuming 200 fs pump. Violet solid line -  PDC spectra, red shaded area -  LO spectra. Orange and green dashed lines represent the intensity distributions of the $HG_{1}$ and $HG_{2}$ PDC modes, respectively.}
\label{fig: Theoretical fs}
\end{figure}

Then, using the broadened LO, we performed squeezing measurements with and without the compressor for both PDC types and compared them with the measurements based on the reference unbroadened LO, the results are presented in Fig. \ref{fig: Squeezing} and Table \ref{table1}.  In these measurements, we considered two spectral profiles of the broadened LO referred to as spectrum 1 and spectrum 2, see Fig. \ref{fig: Spectra LO}. As can be seen from the presented results, the use of the broadened LO without a compressor significantly reduces the degree of squeezing and anti-squeezing for both PDC types compared to the reference LO, so that in the case of type-II PDC, the squeezing and anti-squeezing are not observed at all. Implementing the chirp compensation procedure helps restore the degree of squeezing and anti-squeezing. For example, in the case of type-0 PDC, the levels of squeezing and anti-squeezing were restored to -0.42 dB and 0.91 dB, respectively, for the broadened LO spectrum 1 and, to -0.19 dB and 0.64 dB, respectively, for the broadened LO  spectrum 2. Similarly, using a chirp compensation, we recovered the squeezing and anti-squeezing values for type-II PDC, achieving  -0.19 dB and  0.24 dB for the squeezing and anti-squeezing, respectively,  for the broadened LO  spectrum 1, and -0.11 dB and  0.22 dB, respectively,  for the broadened LO  spectrum 2.  
Further use of the pulse-shaping technique  might help to improve the degree of squeezing in the corresponding measurements.

\begin{table}
\caption{\label{table1} Experimentally measured values of squeezing (Sq) and anti-squeezing (ASq) for the chirped and compensated broadened LO spectrum 1 and spectrum 2 compared with the reference LO for type-0 and type-II PDC.}
\vspace{2mm}
\centerline{
\begin{tabular}{c|c|c|c|c}
\hline
$PDC$ & \multicolumn{2}{c|}{Type-0} & \multicolumn{2}{c}{Type-II} \\
\hline
& Sq, dB & ASq, dB & Sq, dB& ASq, dB \\
\hline
\text{Reference} & -0.45 & 1.04 & -0.31 & 0.44 \\
\hline
\text{Chirped spectrum 1} & -0.11 & 0.15 & -- & -- \\
\text{Chirped spectrum 2} & -0.08 & 0.12 & -- & -- \\
\hline
\text{Comp. spectrum 1} & -0.42 & 0.91 & -0.19 & 0.24 \\
\text{Comp. spectrum 2} & -0.19 & 0.64 & -0.11 & 0.22 \\
\hline
\end{tabular}}
\end{table}

It should be noted that the presented method of the LO broadening is a uniquely effective way to measure  squeezing of the Schmidt modes correctly.  For example, other methods, such as using a shorter pulse, do not improve the matching between the LO and the Schmidt modes. Indeed, although femtosecond pumping provides a broadband LO spanning a range from a few to several tens of nanometers, it simultaneously relaxes the phase-matching conditions for the PDC process, leading to significant broadening of the PDC spectra and Schmidt modes, which is demonstrated in Fig. \ref{fig: Theoretical fs}. Here, we performed numerical simulations according to Eq. (\ref{eq: JSA6} - \ref{eq: Theor spectra}) using the same experimental parameters as described above with the only difference that the pump pulse duration was changed from 2 ps to 200 fs.
It can be easily seen that even though the LO bandwidth constitutes now 3.5 nm, it does not match the modes of both type-0 and type-II PDC,  as their spectral widths also increase. This turns into an endless game in which the use of increasingly short pump pulses leads to the broadening of both the LO and the Schmidt modes, but fails to ensure their matching.
At the same time, it is worth noting that the broadening of the femtosecond LO in microstructured fibers allows increasing its bandwidth in several times while keeping the shape without oscillations.  
To illustrate this, we estimated the uniform broadening of the LO up to 16 nm in the same fiber considering the 200 fs pump (see Appendix \ref{app: Pulse-propagation}). Such bandwidth is comparable to that of the multimode squeezed light produced in other setups discussed in \cite{10.1063/5.0156331, NatPhoton.8.109}.

\section{Conclusions}
In this work, we demonstrated a new technique, which increases the spectral bandwidth of the local oscillator to ensure its matching with the modes of the multimode squeezed light. 
This technique is based on the passing of the LO through the microstructured fiber, where it experienced broadening due to self-phase modulation. The chirp induced by the LO propagation through the fiber  was compensated by employing a pulse compressor. We validated our approach by generating multimode squeezed light in single-pass OPA schemes for both type-0 and type-II PDC and measuring squeezing with the use of the broadened LO. For the case of type-II PDC the broadened LO spectrally matched the Schmidt modes of the generated squeezed light.
 We also demonstrated that the simple approach of increasing the bandwidth of the LO by using shorter pulse durations does not solve the problem of matching, because the bandwidth of the PDC spectrum and the modes are also broadened due to the relaxation of phase-matching conditions.
The proposed scheme is applicable to measuring squeezing in multimode systems with pulsed pumping and is sufficiently flexible, allowing for the adjustment of a number of parameters, such as fiber length, zero-dispersion wavelength, pump power, and polarization. The approach demonstrated in this work may be useful for various CV quantum computing and communications protocols.

\begin{acknowledgments}
We acknowledge the financial support of the RSF grant No. 26-12-00198 and greatly thank Prof. A.V. Masalov, Prof. I.A. Bilenko, Dr. D.A. Chermoshentsev and the members of the Hybrid Quantum Photonics group of RQC  for discussions.
\end{acknowledgments}

\appendix

\section{\label{app: Pulse-propagation}Pulse propagation in the microstructured fiber}

The fiber used was produced by the standard stack and draw method, with the core made of fused silica and air-holes in the cladding region. The propagation of the laser pulse (LO) in the microstructured fiber is considered using the numerical solution of the generalized nonlinear Schrodinger equation \cite{G-P-Agrawal}:
\begin{eqnarray}
\frac{\partial A(z, \tau)}{\partial z} & = & i \sum_{k=2}^{6} \frac{(i)^k}{k!} \beta^{(k)} \frac{\partial^k A}{\partial \tau^k} \nonumber \\ 
&& + i \gamma \left( 1 + \frac{i}{\omega_0} \frac{\partial}{\partial \tau} \right)  
\label{eq:beta}
\\ 
&& \times\left[ A(z, \tau) \int R(\eta) \times |A(z, \tau - \eta)|^2 \, d\eta \right] \nonumber,
\end{eqnarray}
where $A$ is the field amplitude, $\beta^{(k)} = \frac{\partial^k \beta}{\partial \omega^k}$ are the coefficients in the Taylor-series expansion of the propagation constant $\beta(\omega)$ for the fundamental mode, $\omega_{0}$ is the carrier frequency, $\tau$ is the retarded time, $\gamma$ is the nonlinear coefficient, and $R(t)$ is the retarded nonlinear response function given by \cite{Blow_89}:
\begin{eqnarray}
R(t) & = & (1 - f_R) \delta(t) \nonumber \\ 
&& + f_R \Theta(t) \frac{\tau_1^2 + \tau_2^2}{\tau_1^2 \tau_2^2} e^{-t/\tau_2} \sin\left(\frac{t}{\tau_1}\right),
\label{eq: response}
\end{eqnarray} 
where $f_R=0.18$ is the fractional contribution of the Raman response; $\delta(t)$ and $\Theta(t)$ are the delta and the Heaviside step functions, respectively; $\tau_1=12.5 \ fs$ and $\tau_2=32 \ fs$ are the characteristic times of the Raman response of fused silica.
We also consider nonlinear coefficient as $\gamma=(n_2\omega_0)/(cS_{eff})$ , where $n_2$ is the nonlinear refractive index (Kerr nonlinearity) of the microstructured fiber material, which equals for the fused silica to $n_2=3.2\times10^{-16} \ cm^2/W$, and 
\begin{equation}
    S_{eff}=\frac{(\int_{-\infty}^{\infty}\int_{-\infty}^{\infty}|F(x,y)^2dxdy|)^2}{\int\int|F(x,y)^4dxdy|}
\end{equation}
is the effective mode area, where $F(x,y)$ is the fundamental mode transverse field profile. 

\begin{figure}[h]
\centering\includegraphics[width=0.4\textwidth]{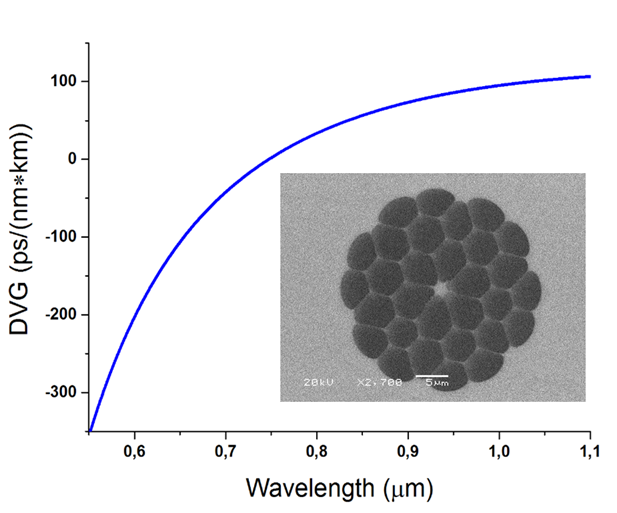}
\caption{Group-velocity dispersion (GVD) as a function of the wavelength for the fundamental mode of a fused silica microstructured fiber. At the inset: the SEM image of the fiber.}
\label{fig: GVD}
\end{figure}

The parameters of the fiber, $\beta^{(k)}$ and $\gamma$, were found by applying an algorithm that uses a polar-coordinate Fourier decomposition method with adjustable boundary conditions within the scalar wave approximation, developed by Poladian et al \cite{Poladian:02}.  The frequency dependence of the propagation constant $\beta(w)$ for the fundamental mode of microstructured fiber computed with the use of this numerical procedure yields the following $\beta^{(k)}$ coefficients for the central wavelength at 783.3 nm: $\beta^{(2)}=-0.0155 \ ps^2/m$, $\beta^{(3)}=7.13\times10^{-5} \ ps^3/m$, $\beta^{(4)}=-3.65\times10^{-8} \ ps^4/m$, $\beta^{(5)}=0.77\times10^{-10} \ ps^5/m$, $\beta^{(6)}=-5.04\times10^{-6} \ ps^6/m$. The blue curve at Fig. \ref{fig: GVD} displays the $GVD=-2\pi c \lambda^{-2} \beta^{(2)}(\lambda)$ for the fundamental mode of the microstructured fiber as a function of the wavelength $\lambda$. The corresponding nonlinear coefficient at 783.3 nm was estimated as 80 $(W \ km^{-1})$.

\begin{figure*}[t]
\centering\includegraphics[width=0.8\textwidth]{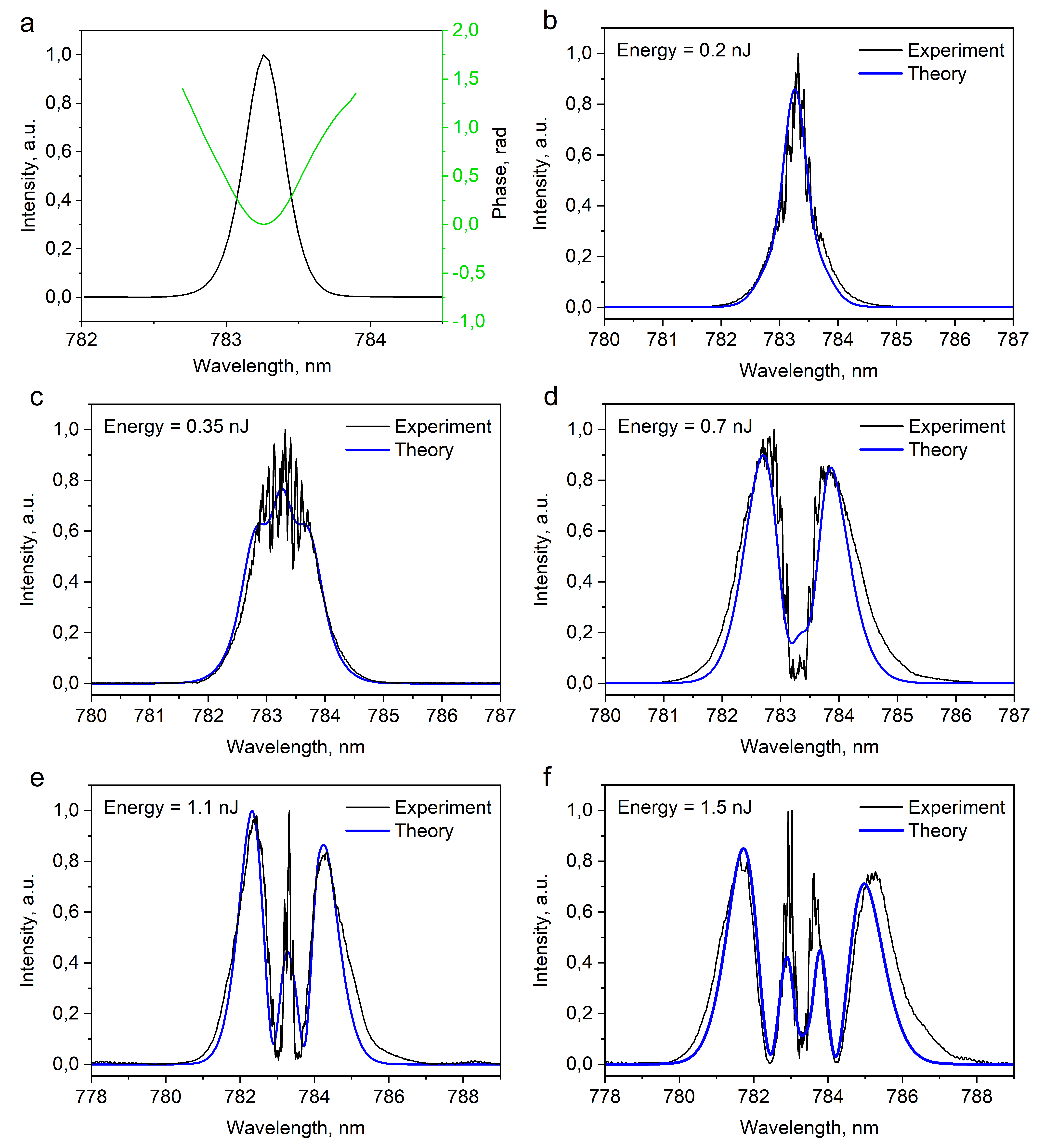}
\caption{Evolution of the pulse in the microstructured fiber: a) intensity spectrum (black line) and spectral phase (green line) of the input laser pulse with the initial duration of 2 ps. b - f) Experimental (black line) and theoretical (blue line) spectra measured and calculated at the output of a 21-cm microstructured fiber with the initial light energy of (b) 0.2 nJ, (c) 0.35 nJ, (d) 0.7 nJ, (e) 1.1 nJ and (f) 1.5 nJ, respectively}
\label{fig: Evolution in the MF}
\end{figure*}

\begin{figure}[htbp]
\centering\includegraphics[width=0.4\textwidth]{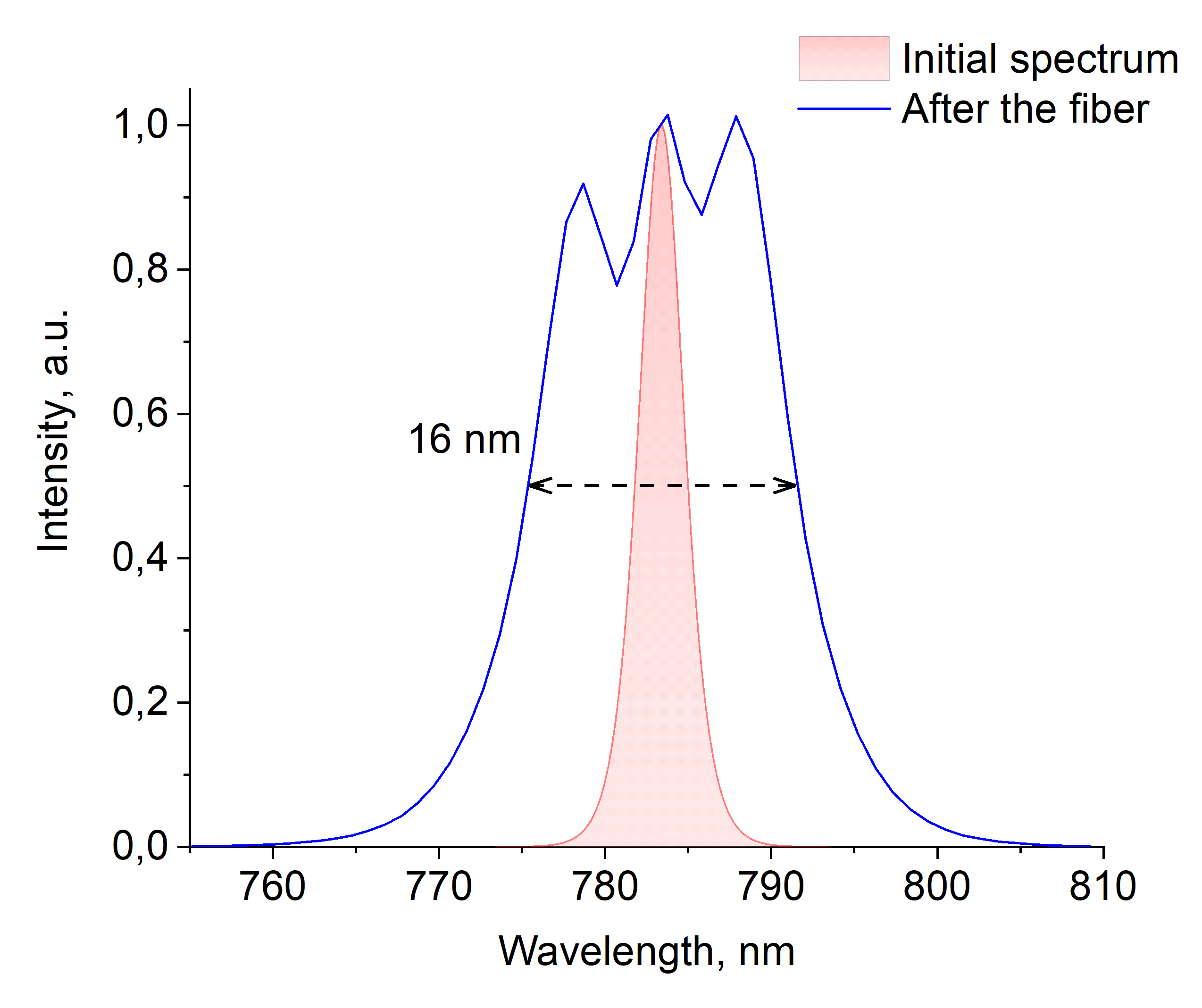}
\caption{Results of calculations for the pulse broadening in the microstructured fiber at 35 pJ. The initial spectrum of 3.5 nm (red shaded area) uniformly broadens  up to 16 nm (blue curve).}
\label{fig: fs-pulse evolution}
\end{figure}

Eq. (\ref{eq:beta}) was applied to compute the evolution of a 2 ps laser pulse after its propagation through 21-cm of microstructured fiber with the cross section shown in the inset of Fig. \ref{fig: GVD}. An initial spectrum and spectral phase for the input laser pulse with the central wavelength 783.3 nm are shown in Fig. \ref{fig: Evolution in the MF}(a). A comparison of the output experimental and theoretical spectra for pulses with different initial energies (0.2 nJ, 0.35 nJ, 0.7 nJ, 1.1 nJ, 1.5 nJ) is presented in Fig. \ref{fig: Evolution in the MF}(b)-(f). The input radiation energy coupled into the fundamental mode of the microstructured fiber was used as a fitting parameter in our simulations and was defined from the best fit between the simulated spectra and the experimental data. Fig. \ref{fig: Evolution in the MF}(b)-(f) shows that spectral broadening is growing as the energy of the input pulse increases from 0.2 to 1.5 nJ. According to Eq. (\ref{eq:beta}) the main contribution to such spectral transformation is made by self-phase modulation induced by the Kerr nonlinearity. The estimated parameters of the fiber allow to predict the evolution of the shorter pulse, i.e. 200 fs. In this case, sending a 35 pJ pulse through the fiber leads to the uniform broadening from 3.5 to 16 nm, see Fig. \ref{fig: fs-pulse evolution}.

\section{\label{app: Chirp-control}Chirp characterization}
\subsection{\label{app: SHG}Second-harmonic generation}

\begin{figure*}
\centering\includegraphics[width=0.8\textwidth]{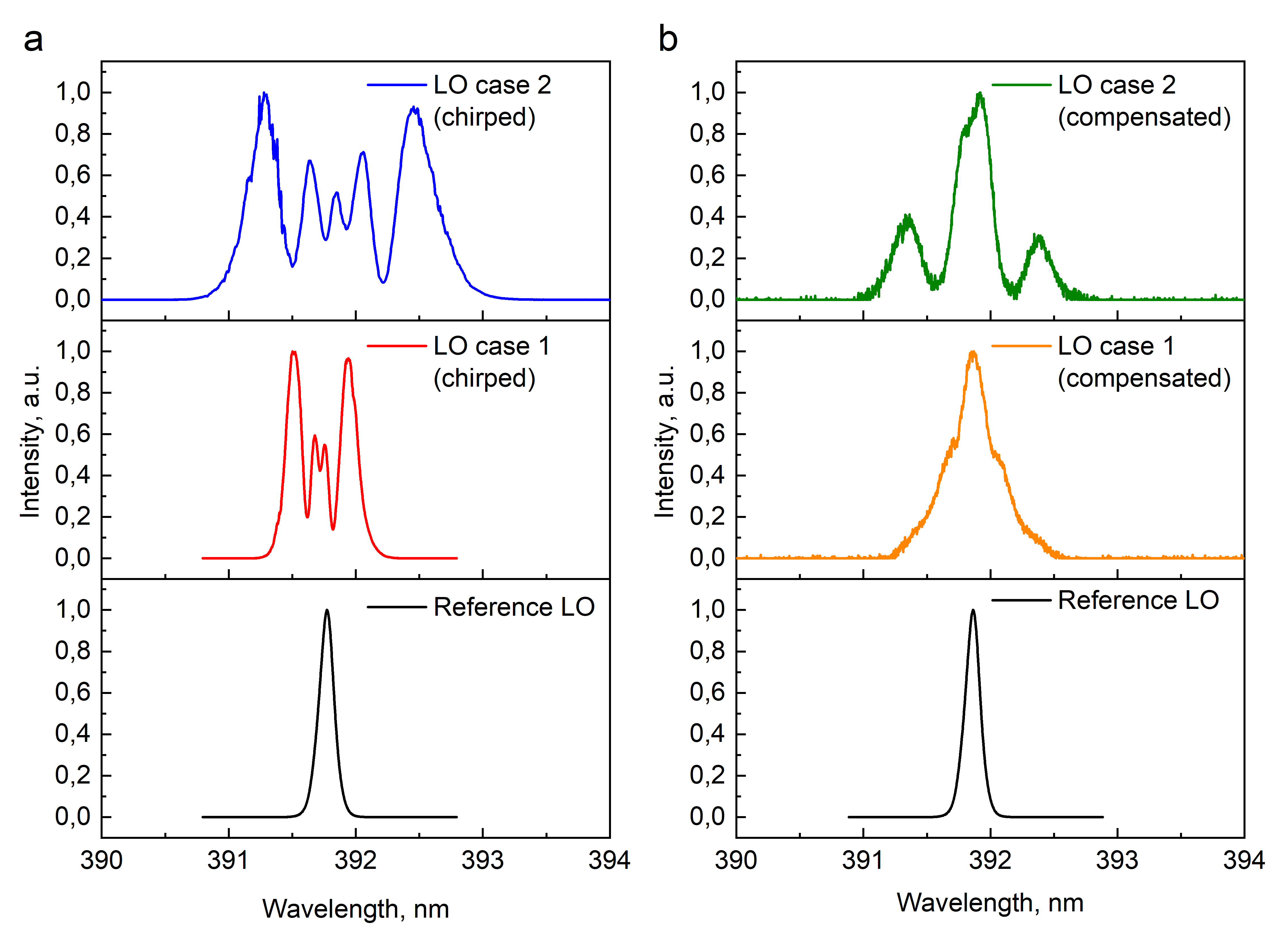}
\caption{Spectra of the second harmonic produced from the broaden LO a) without and b) with the chirp compensation. Case 1 and case 2 correspond to the spectrum 1 and spectrum 2 of the broadened LO in Fig. \ref{fig: Spectra LO}, respectively. The black line indicates the second harmonic spectrum from the reference (unbroaden) LO. }
\label{fig: SHG}
\end{figure*}

We characterized the chirp compensation by measuring the second-harmonic generation. For this, the LO was sent to the 0.5 mm BBO crystal cut for type-I phase matching, the generated second harmonic was then fed into the single-mode fiber and measured by an optical spectral analyzer (Yokogawa-AQ6373B). We measured the second harmonic from the reference LO and the broadened LO (cases 1 and 2)  without (Fig. \ref{fig: SHG}a) and with (Fig. \ref{fig: SHG}b) the chirp compensation. It is clearly seen that the chirp compensation results in a distinctive peak at the central wavelength, when all spectral components of the LO contribute coherently, unlike the case without compensation where multiple peaks are observed.

\subsection{\label{app: PulseDuration}Pulse duration measurement}

\begin{figure*}[t]
\centering\includegraphics[width=0.9\textwidth]{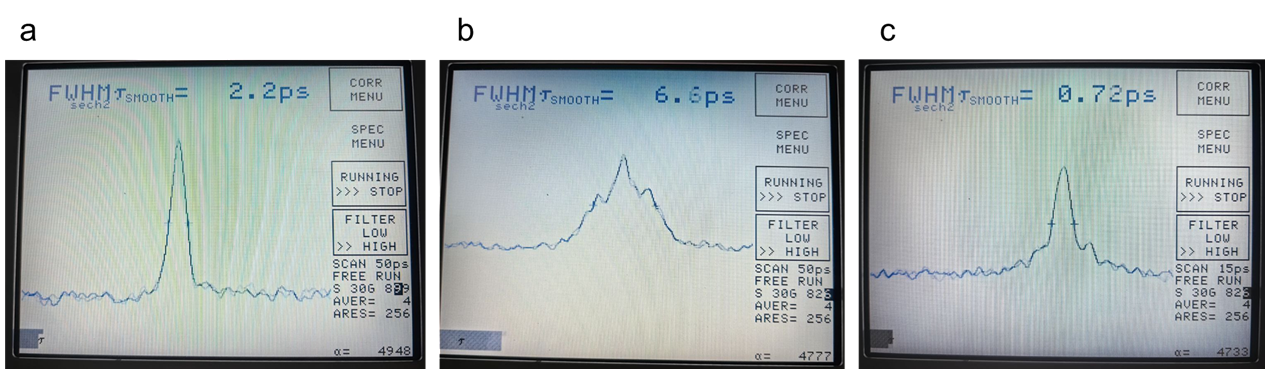}
\caption{Pulse duration of the a) reference (unbroadened) LO and broadened LO after propagating through the fiber  b) without and c) with the chirp compensation.}
\label{fig: Pulse duration measurement}
\end{figure*}

The pulse duration of the LO was measured  by the APE PulseCheck autocorrelator.  Fig. \ref{fig: Pulse duration measurement} shows the results of such measurements: the reference LO (Fig. \ref{fig: Pulse duration measurement}a) was compared with the LO after propagating through the microstructered fiber without (Fig. \ref{fig: Pulse duration measurement}b) and with (Fig. \ref{fig: Pulse duration measurement}c) chirp compensation. It can be seen that the propagation in the fiber leads to the increase in the pulse duration from  2 ps to almost 7 ps due to the delay of short-wave components with respect to the long-wave ones, while implementing the compressor allows us to compensate for temporal broadening and reduce the pulse duration to 0.7 ps.

\subsection{\label{app: X-FROG}X-FROG measurement}
\begin{figure}[h]
\centering\includegraphics[width=0.4\textwidth]{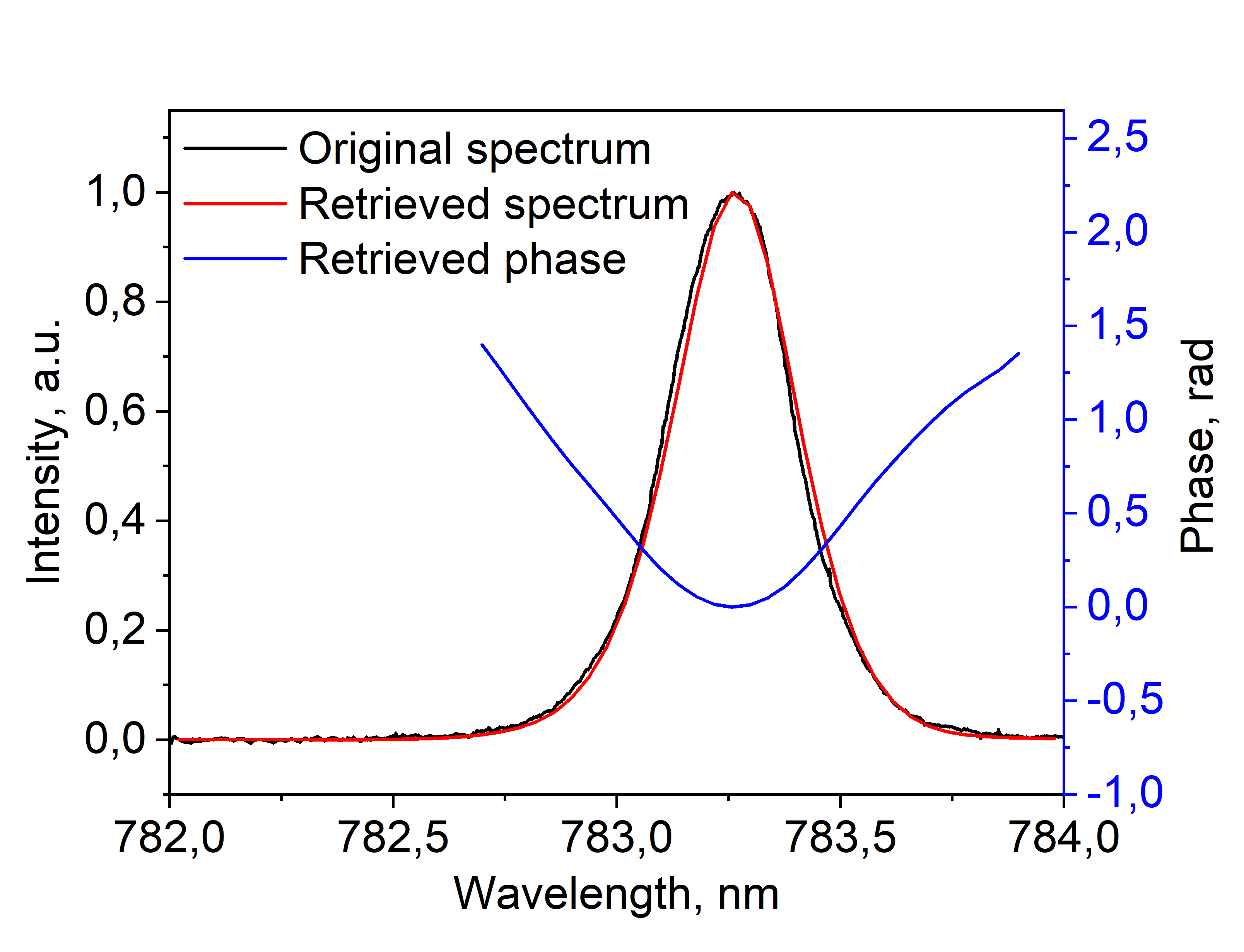}
\caption{Retrieved spectrum (red) and phase (blue) of the reference LO compared to the measured spectrum (black)}
\label{fig: FROG_LO}
\end{figure}

A complete phase characterization of the employed LO was performed using the Cross-correlation Frequency-resolved Optical Gating (X-FROG) technique \cite{trebino2012frequency}. In this technique, the initial LO (the laser beam from the Ti:Sapphire laser) was used as the reference beam. First, we characterized the reference LO to retrieve its phase profile using the FROG technique, where the reference LO is combined with itself. The retrieved results are shown in Figure \ref{fig: FROG_LO}.

\begin{figure*}[htbp]
\centering\includegraphics[width=0.85\textwidth]{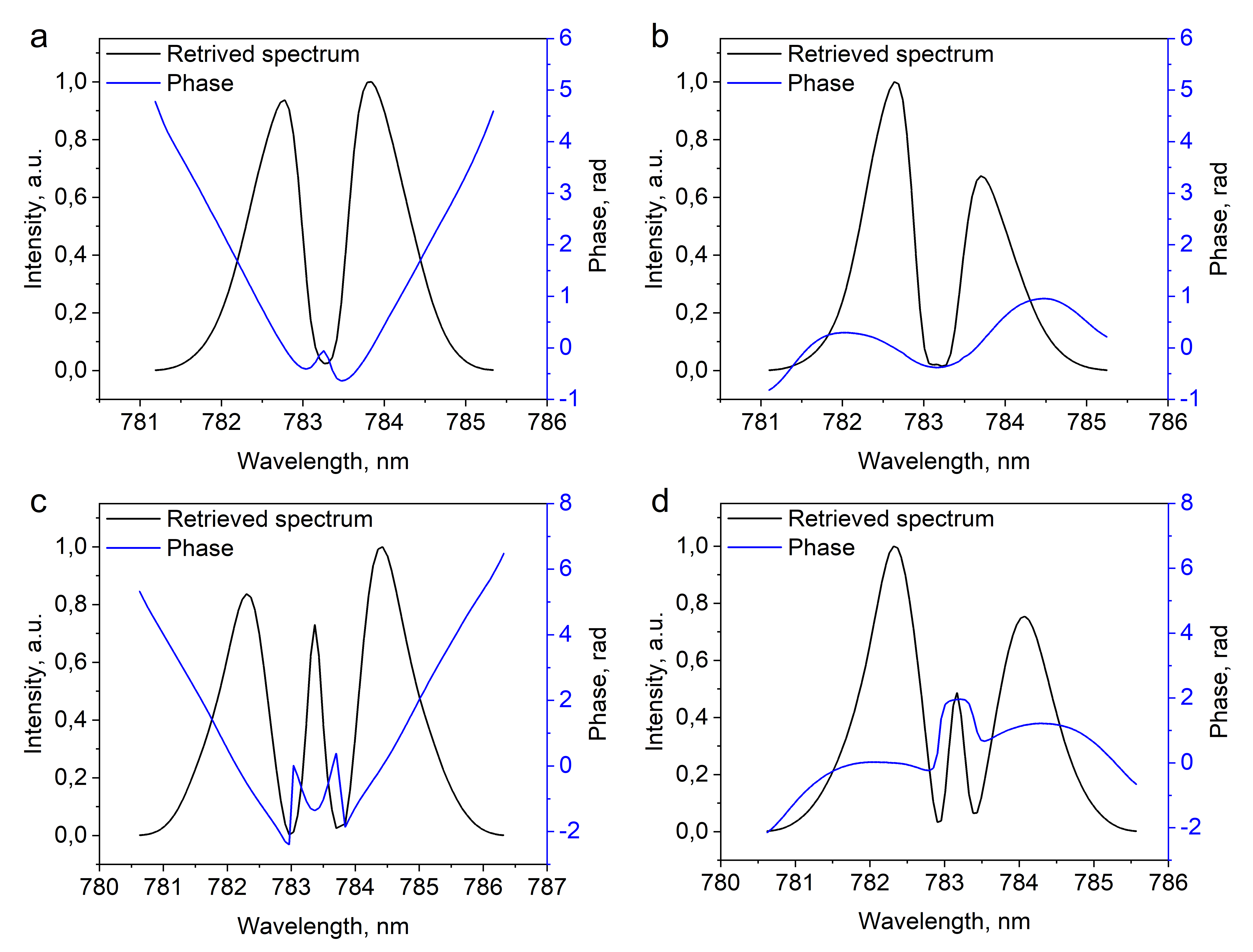}
\caption{Retrieved spectrum (black) and phase (blue) using the X-FROG technique for: a) spectrum 1 without compressor, b) spectrum 1 with compressor, c) spectrum 2 without compressor, d) spectrum 2 with compressor.}
\label{fig: X-FROG}
\end{figure*}

Then we implemented the obtained results in the X-FROG technique, where the broadened LO is recombined with the reference one, to study the LO for the cases 1 and  2 without and with the chirp compensation, the results are presented in Fig. \ref{fig: X-FROG}. It can be seen that for the LO undergoing chirp compensation the phase variation is significantly reduced. We also validated our X-FROG retrieval algorithm  by comparing frequency marginals with the independently measured spectra of the LO. The obtained X-FROG error of about $10^{-3}$ demonstrates a good quality of the retrieval algorithm. In addition, we compared the FWHM of the pulse duration retrieved using the X-FROG method with the one directly measured by the autocorrelator, in both cases it was close to 0.7 ps for the broadened LO after the compressor.

\bibliography{References}

\end{document}